\documentclass[prl,aps,twocolumn,showpacs,superscriptaddress]{revtex4-1}
\usepackage{graphics}
\usepackage{epsfig}
\usepackage[utf8]{inputenc}
\usepackage{graphicx}
\usepackage{amsmath}
\usepackage{amsfonts}
\usepackage{amssymb}
\usepackage{graphicx}
\usepackage{xcolor}
\usepackage{tabularx}
\usepackage{txfonts}
\usepackage{appendix}
\usepackage{xcolor}

\begin{document}
\title{Strain induced electronic and magnetic transition in $S = \frac{3}{2}$ antiferromagnetic spin chain compound LaCrS$_3$}

\author{Kuldeep Kargeti}
\affiliation{Department of Physics, Bennett University, Greater Noida 201310, Uttar Pradesh, India}
\author{Aadit Sen}
\affiliation{Department of Physics, Bennett University, Greater Noida 201310, Uttar Pradesh, India}
\author{ S. K. Panda}
\email{swarup.panda@bennett.edu.in}
\affiliation{Department of Physics, Bennett University, Greater Noida 201310, Uttar Pradesh, India}

\begin{abstract}
Exploring the physics of low-dimensional spin systems and their pressure-driven electronic and magnetic transitions are thriving research field in modern condensed matter physics. In this context, recently antiferromagnetic Cr-based compounds such as CrI$_3$, CrBr$_3$, CrGeTe$_3$ have been investigated experimentally and theoretically for their possible spintronics applications. Motivated by the fundamental and industrial importance of these materials, we theoretically studied the electronic and magnetic properties of a relatively less explored Cr-based chalcogenide, namely LaCrS$_3$ where 2D layers of magnetic Cr$^{3+}$ ions form a rectangular lattice. We employed density functional theory + Hubbard $U$ approach in conjunction with constrained random-phase approximation (cRPA) where the later was used to estimate the strength of $U$.  Our findings at ambient pressure show that the system exhibits semiconducting antiferromagnetic ground state with a gap of 0.5 eV and large Cr moments that corresponds to nominal S=3/2 spin-state. The 1st nearest neighbor (NN) interatomic exchange coupling (J$_1$) is found to be strongly antiferromagnetic (AFM), while 2nd NN couplings are relatively weaker ferromagnetic (FM), making this system a candidate for 1D non-frustrated antiferromagnetic spin-chain family of materials. Based on orbital resolved interactions, we demonstrated the reason behind two different types of interactions among 1st and 2nd NN despite their very similar bond lengths. We observe a significant spin-orbit coupling effect, giving rise to a finite magneto crystalline anisotropy, and Dzyaloshinskii-Moriya (DM) interaction. Further, we found that by applying uniaxial tensile strain along crystallographic $a$ and $b$-axis, LaCrS$_3$ exhibits a magnetic transition to a semi-conducting FM ground state, while compression gives rise to the realization of novel gapless semiconducting antiferromagnetic ground state. Thus, our findings can enrich the versatility of LaCrS$_3$ and make it a promising candidate for industrial applications.
\end{abstract}

\maketitle

\section{Introduction}
The electronic structure and magnetism of one-dimensional (1D) spin chain systems is a thriving research field in modern condensed matter physics~\cite{frustratedmagnetism-book} as they are used for the analyses of many fundamental concepts in many-body quantum physics as well as have great technological promises for quantum computation and information processing. The quantum spin fluctuations are an intrinsic characteristic of such low-dimensional magnetic systems having localized spin moments. These fluctuations, the rich electron correlated behavior, and spin-orbit coupling often lead to emergent phenomena such as Haldane gap states ~\cite{haldane, Kenzelmann}, new topological phases of matter~\cite{Burch2018}. In this direction, the Cr-based semiconducting materials with layered structures have recently obtained substantial attention due to its potential application for many technologies from sensing to data storage. For instance, the experimental demonstration of long range magnetism in layered materials, e.g. Cr trihalides (CrX$_3$, X = I, Br and Cl) is considered as a major breakthrough in the field of low dimensional magnetism since it opens up opportunities for sprintronics applications~\cite{wang2011electronic}. There are also reports of strain-based tuning of magnetism in layered materials such as CrGeTe$_3$ which has emerged as an excellent magnetic substrate in nanoelectronic devices and also for next-generation memory devices~\cite{shang2019stacking}. 
\par
\begin{figure}[t]
\includegraphics[width=0.99\columnwidth]{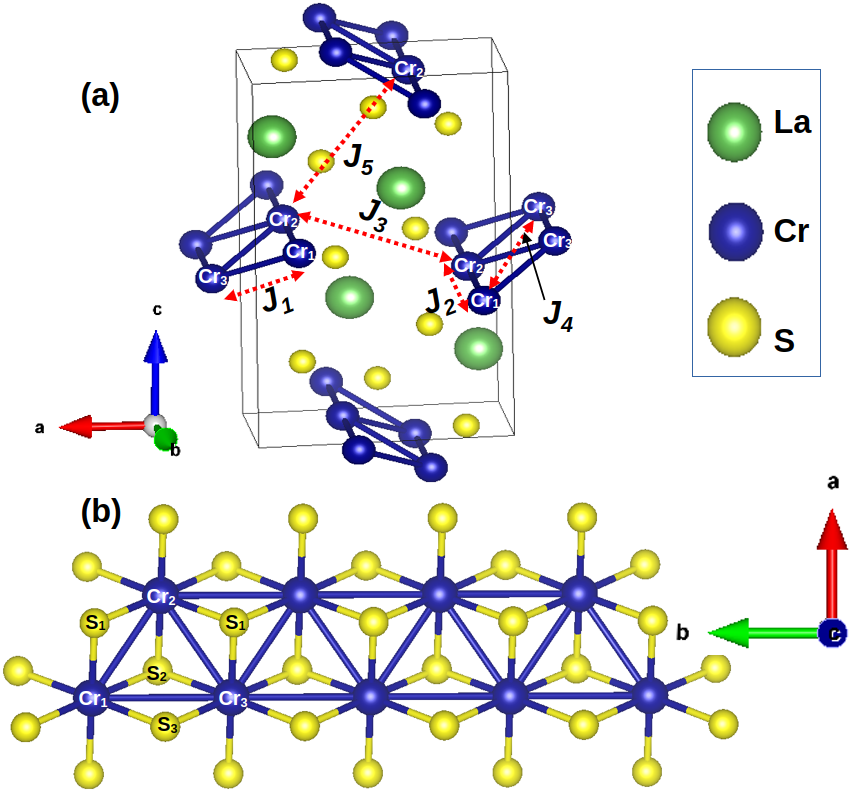}
\caption{(a) The unit cell of LaCrS$_3$. The magnetic couplings between nearest neighbor Cr-ions are marked. (b) The CrS$_6$ edge-sharing octahedral
units form the zigzag chain running along the $b$-axis.}
\label{cryststruct}
\end{figure}

 
In view of the fact that layered Cr-based materials have huge fundamental and industrial importance, we theoretically studied the electronic and magnetic properties of a relatively less explored Cr-based chalcogenide, namely LaCrS$_3$. The crystal structure of LaCrS$_ 3$ is displayed in Fig.~\ref{cryststruct}(a) where 2D layers of magnetic Cr$^{3+}$ ions form a rectangular lattice. The edge sharing CrS$_6$ unit forms a double chain along $b$-direction and each Cr is connected to other two Cr ions via S-ions as illustrated in Fig.~\ref{cryststruct}(b). The Cr-Cr distances in the triangles of this double chain are 3.47~\AA ~(Cr$_1$-Cr$_2$ and Cr$_2$-Cr$_3$) and 3.83~\AA (Cr$_1$-Cr$_3$), respectively and the inter-chain Cr-Cr distance in the $ab$-plane (the third nearest neighbor: Cr$_2$-Cr$_2$ in Fig.~\ref{cryststruct}(a)) is 5.83 ~\AA. The Cr-S bond distances of CrS$_6$ octahedra lie in the range between 2.38~\AA and  2.48~\AA. The experimental resistivity data~\cite{KIKKAWA1998233} indicates that the ground state is insulating. The magnetic susceptibility ($\chi$)~\cite{KIKKAWA1998233} increases with decreasing temperature and passes through a broad maximum at around 195 K which is a typical feature of short-range spin correlation \cite{spin-correlation}. The high-temperature fitting of $\chi$ gives the Weiss temperature to be -8.5 K, indicating antiferromagnetic coupling between the Cr-ions. At lower temperatures well below the broad maximum, the $\chi$ decays exponentially which could be attributed to the opening of a spin gap. 

\par 
The microscopic understanding of the magnetism of LaCrS$_3$, particularly to find out if it belongs to the family of spin-chain compounds requires a detailed knowledge of the magnetic spin Hamiltonian. Such understanding will also help us to predict the efficient ways of tuning its spin interactions, magnetic, electronic phases and emergence of novel magnetic phenomena~\cite{PhysRevB.97.020402} which are important for technological and scientific aspects. The tuning of the various properties such as modification of inter-atomic exchange couplings, magnetic anisotropies, critical magnetic transition temperature, band-gap tuning etc are possible under the influence of mechanical strain. To the best of our knowledge, no theoretical calculations have been done so far to comprehend the electronic structure and magnetism of this intriguing correlated material that belongs to the transition metal chalcogenide family. 
\par 
In this work, we have investigated the electronic and magnetic properties of LaCrS$_3$ using density functional theory + Hubbard $U$ method under ambient conditions and also observed how its properties could be tuned under the influence of compressive and tensile uniaxial strain. Our results show that the 1st nearest neighbor (NN) and 2nd NN spin interaction strengths between Cr ions are crucially different despite their very similar bond length. In fact, 1st NN interactions are strong antiferromagnetic, while 2nd NN interactions are relatively weaker FM, making this system a 1D non-frustrated antiferromagnetic S=3/2 spin-chain compound at ambient conditions. We found a non-negligible spin-orbit coupling effects, giving rise to a finite MAE, and DM interactions. Further, we proposed ways to realize magnetic transition (AFM to FM), normal semiconductor to a gapless semiconductor transition via application of uniaxial strain. 

\begin{figure}[t]
\includegraphics[width=0.99\columnwidth]{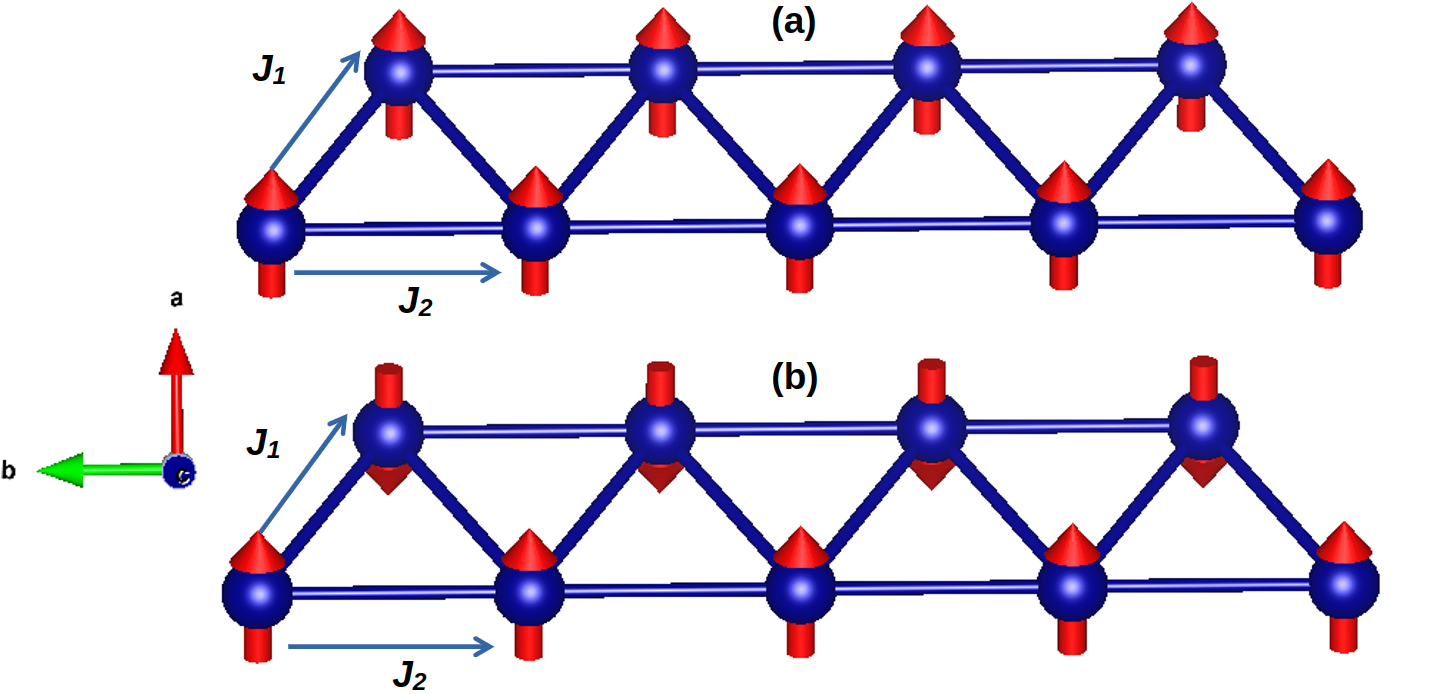}
\caption{(a) The ferromagnetic ordering, and (b) antiferromagnetic ordering between the nearest neighbor Cr-ions. These two magnetic  configurations are considered to get the lowest energy state.}
\label{configurations}
\end{figure}

\section{Computational details}
All calculations reported in this work are carried out using three approaches, namely (a) a plane-wave basis as implemented in Vienna ab initio simulation package (VASP)~\cite{vasp1,vasp2} with projector augmented wave potentials~\cite{paw}, (b) the full potential linearized augmented plane wave (FP-LAPW) basis as implemented in WIEN2K code~\cite{wien2k} and full-potential linear muffin-tin orbital (FP-LMTO) method~\cite{FPLMTO_Orig, FPLMTO} as implemented in the RSPt code~\cite{FPLMTOCode}. The calculations have been cross-checked within these three sets of methods providing further credence to our obtained results. Exchange and correlation effects are treated using generalized gradient approximation (GGA)~\cite{GGA_PBE} as well as including Hubbard $U$ within the GGA+U framework~\cite{LSDAU_Liechtenstein}. In order to avoid any ambiguity about the choice of $U$ and to make the GGA+U approach completely parameter-free, we computed the values of effective Coulomb interaction ($U$) within the framework of constrained random-phase approximation (cRPA)~\cite{cRPA_orig} as implemented in the WIEN2K code~\cite{cRPA_wien2kImple}. The results of our calculations from a so called $d$-$d$ model~\cite{panda-NiO} show that this method yielded a $U$ = 1.6 eV and average Hund's coupling $J$ = 0.5 eV for the $3d$ states of Cr in LaCrS$_3$. A $6\times 12\times 4$  $k$-mesh has been used for the Brillouin zone (BZ) integration. 
\par 
Using the converged GGA+U solutions of RSPt~\cite{FPLMTOCode}, we employed the magnetic force theorem~\cite{magneticforceth1,magneticforceth2} to extract the effective inter-site exchange parameters ($J_{ij}$). A detailed discussion of the implementation of the magnetic force theorem in RSPt is provided in Ref.~\onlinecite{PhysRevB.91.125133}. The effective $J_{ij}$ is extracted in a linear-response manner via a Green's function technique. We have successfully used this method for other transition metal compounds~\cite{PhysRevB.94.064427}.

\begin{figure*}[t]
\includegraphics[width=1.99\columnwidth]{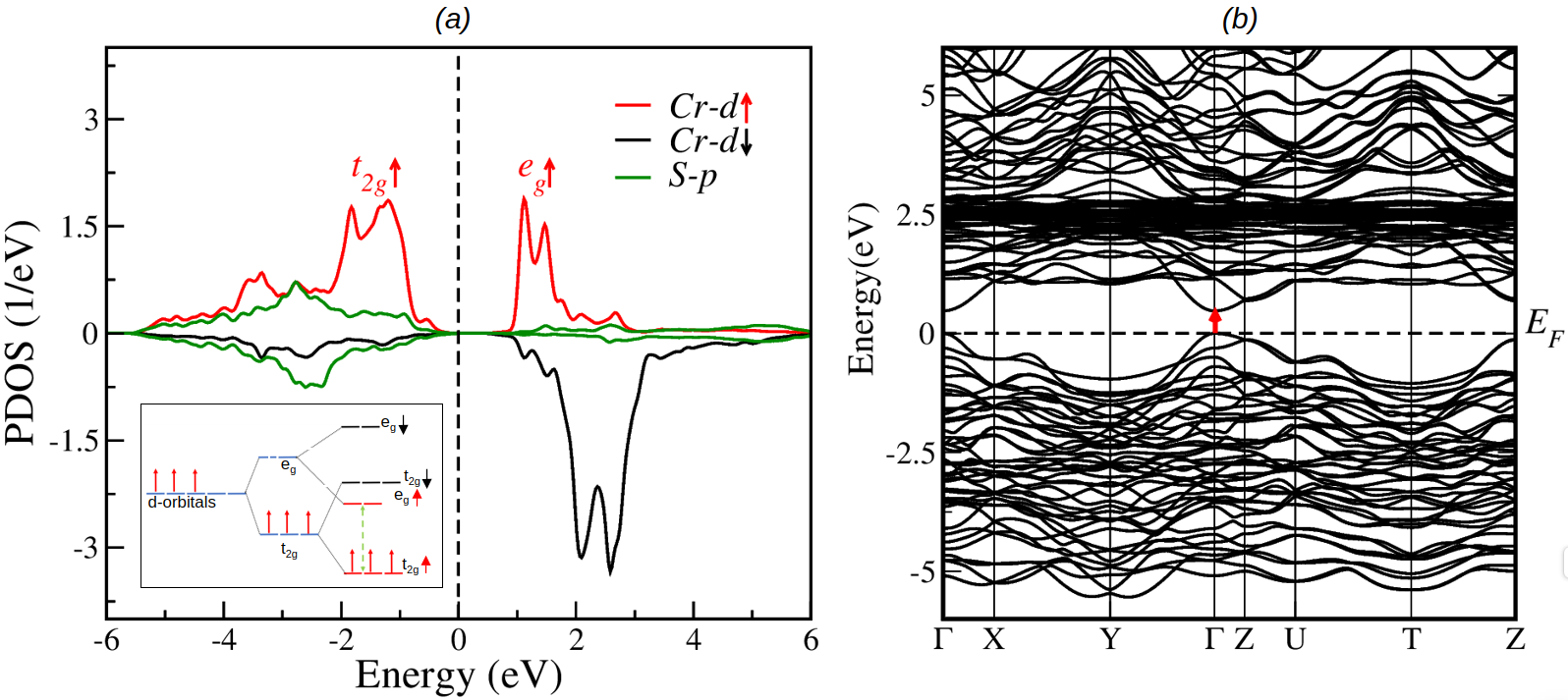}
\caption{(a) The computed spin-polarized partial density of states of $Cr-d$, and $S-p$ states. The Fermi energy has been set at 0 eV. In the inset, a schematic has been shown to explain the filling of $d$-orbitals as obtained in our calculations and realization of $S=\frac{3}{2}$ state. (b) Band dispersion of the AFM state along various high symmetry directions. The direct gap at $\Gamma$-point has been marked.}
\label{pdosmag}
\end{figure*}

\bgroup
\def\arraystretch{1.0}
\begin{table}[t]
\caption{The total energy (meV/formula unit) of the AFM state with respect to the FM state is shown for both GGA and GGA+U approaches. The magnetic moment of each Cr ion is also listed.}
\vspace{0.1cm}
\begin{tabular}{|   c    | c  c  | c  c  | }
\hline
 & \multicolumn{2}{c|}{GGA} & \multicolumn{2}{c|}{GGA+U} \\
  & \hspace{0.02cm} Energy ($meV$) \hspace{0.02cm} & \hspace{0.02cm} Moment ($\mu_\text{B}$)  \hspace{0.02cm} &  \hspace{0.02cm} Energy ($meV$) \hspace{0.02cm} & \hspace{0.02cm} Moment ($\mu_\text{B}$)  \hspace{0.02cm} \\
 \hline
 FM & 0.00 & 2.90 & 0.00 & 2.90  \\
\hline
 AFM & -23.50 & 2.83 & -14.53 & 2.89 \\
\hline
\end{tabular}
\label{energy-table}
\end{table}
\egroup

\section{Results and Discussion}
\subsection{Magnetism and Electronic structure at ambient condition}
We start with the analysis of magnetic ground state and electronic structure for the bulk LaCrS$_3$ at ambient conditions using various levels of approximations in DFT such as GGA~\cite{PhysRevB.21.5469} and GGA+U~\cite{PhysRevB.44.943}. From the experimental susceptibility data, we know that the system is antiferromagnetic in nature. To obtain an understanding of magnetism, we carried out total energy calculations for two possible magnetic states as schematically presented in Fig.\ref{configurations} . The results of our calculations as demonstrated in Table.~\ref{energy-table} reveal that antiferromagnetic ordering between the NN Cr ions is energetically favorable which is consistent with experiment. The magnitude of this energy difference ($\Delta$E = E$_{FM}$-E$_{AFM}$) becomes slightly smaller in GGA+U scheme compared to GGA. This is not surprising as inclusion of on-site Coulomb repulsion $U$ for the $3d$ electrons of Cr-ions make the $d$ orbitals more localized and as a consequence, inter-atomic exchange coupling is reduced.  Since the value of Hubbard $U$ for Cr-$3d$ states is estimated using cRPA, all calculations in the remainder of the paper have been carried out using GGA+U method.
\par 
We have now analyzed the electronic structure of the antiferromagnetic state in Fig.\ref{pdosmag}.  From crystal geometry, we can see that Cr atoms are surrounded by the octahedral environment formed by sulfur atoms. This leads to the splitting of $d$-orbitals into $t_{2g}$ and $e_g$ states (see schematic in the inset of Fig.\ref{pdosmag}(a)). The computed partial DOS in Fig. \ref{pdosmag}(a) shows that Cr $t_{2g}$ majority spin-channel is fully occupied. The minority $t_{2g}$ and $e_g$ states appear above the Fermi level. This arrangement of $d$-orbitals is consistent with the nominal $d^3$ configuration and $S = \frac{3}{2}$ spin-state for Cr ions according to the Hund's rule as schematically demonstrated in the inset of Fig.\ref{pdosmag}(a). The dominant spectral weight of S-$p$ states arise around 3 eV below the Fermi level. However, there is a small S-$p$ spectral weight in the region where Cr majority $t_{2g}$ states are seen, indicating a weak Cr-S hybridization due to the pi-bonding between Cr-$t_{2g}$ and S-$p$ orbitals.  Fig.\ref{pdosmag}(b) represents the band-dispersion along the high symmetry directions. It reveals that a band-gap of 0.5 eV appears at $\Gamma$-point and hence this system belongs to a direct-gap semiconductor. The gap is seen between the majority and $t_{2g}$ and $e_{g}$ states (Fig.\ref{pdosmag} (a)) and the origin of this gap could be attributed to the combined effects of crystal field, local exchange splitting in Cr-site and electronic correlation of localized $d$-states. The semiconducting ground state as predicted from our GGA+U calculations is in agreement with the experimental data \cite{KIKKAWA1998233}. From the analysis of the electronic structure, we also understood that the half filled $t_{2g}$ (nominal d$^3$ configuration) states are responsible for the magnetism in this system and the computed spin moment at Cr-site turns out to be 2.89 $\mu_B$. Such a high value of moment again emphasizes the localized nature of the Cr-$d$ orbitals. 
\begin{figure}
\includegraphics[width=0.99\columnwidth]{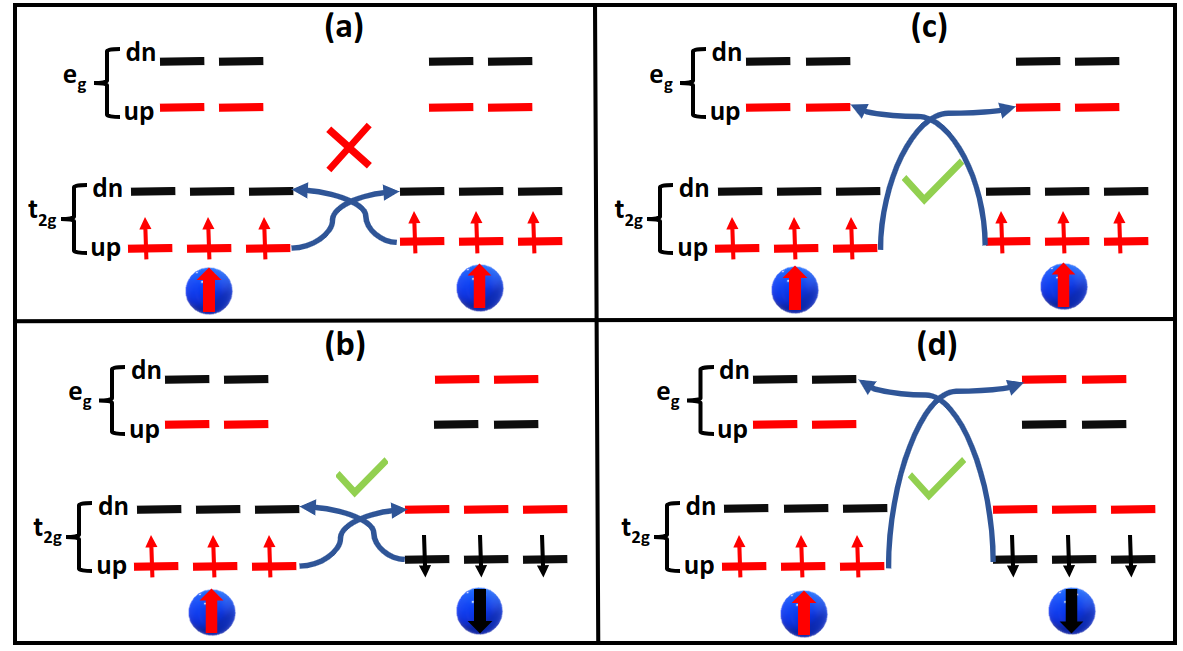}
\caption{The hopping being spin-conserved, the schematic demonstrate that intersite hopping between $t_{2g}$-states are (a) forbidden when Cr-ions are ferromagnetically aligned, (b) and allowed  when they are antiferromagnetically aligned. Similarly (c), (d) $t_{2g}$-$e_g$ intersite hopping is allowed for both kind of coupling. However, ferromagnetic coupling is energetically favored due to Hund's coupling}
\label{schematic}
\end{figure}
\par 
In order to understand the magnetism in detail, we have computed the nature and strength of inter-atomic Heisenberg exchange interaction between the neighboring Cr-spins by employing the magnetic force theorem~\cite{magneticforceth1,magneticforceth2}. We have estimated the strengths of total five interactions where $J_1$, $J_2$ and $J_4$ are intra-chain, while $J_3$ and $J_5$ represent the inter-chain coupling as marked in Fig.~\ref{cryststruct}(a). The computed inter-site exchange interactions corresponding to various Cr-Cr bond lengths are $J_1$ = -7.5 meV, $J_2$ = 4.1 meV, $J_3$ = -0.4 meV, $J_4$ = 0.1 meV, and $J_5$ = -0.4 meV, where negative sign indicates antiferromagnetic and positive sign corresponds to ferromagnetic coupling.  We can see that magnitude of $J_1$ is 1.8 times stronger than $J_2$. This shows a dominant antiferromagnetic exchange between the Cr ions which compliments the experimental findings of kikkawa \textit{et.al.} \cite{KIKKAWA1998233}. The almost negligible magnitudes of $J_3$ and $J_5$ reveal that the inter-chain couplings along both $a$- and $c$-directions are very weak and the magnetism of this system could be analyzed in the form of a spin chain which proceeds along $b$ direction (Fig.~\ref{configurations}(b)). A strong antiferromagnetic $J_1$, relatively weaker ferromagnetic $J_2$, and a high value of spin moment (close to 3 $\mu_B$) make this an ideal candidate for $S = \frac{3}{2}$ antiferromagnetic spin chain family of materials.
\par 
\begin{figure}
\includegraphics[width=0.99\columnwidth]{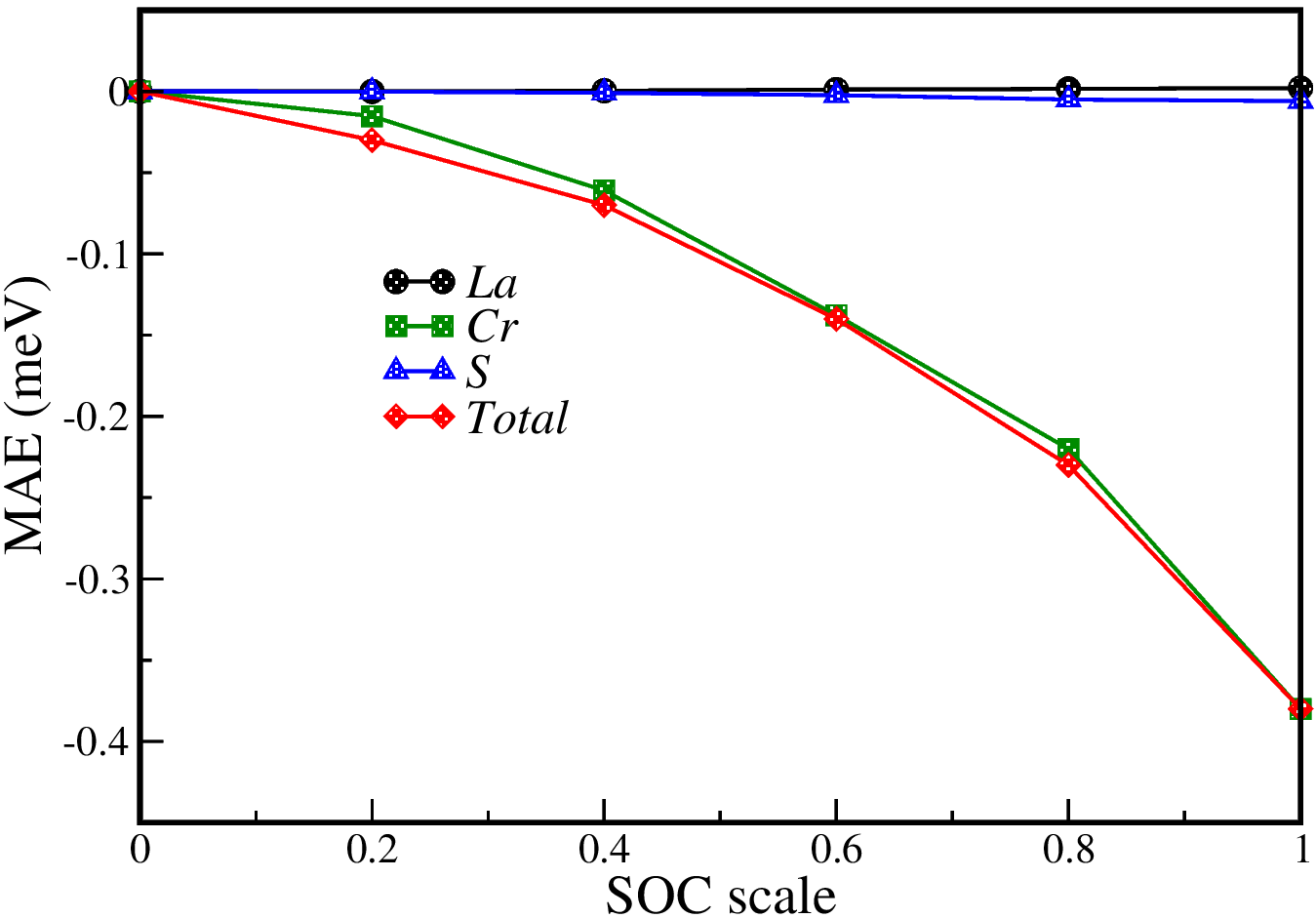}
\caption{The contributions of La, Cr and S atoms to the magneto-crystalline anisotropy as a function of SOC strength. Note that the value of 1.0 represent the SOC strength that corresponds to the actual strength and 0 implies that there is no SOC }
\label{SOC}
\end{figure}

\bgroup
\def\arraystretch{1.5}
\begin{table*}[t]
\caption{Orbital resolved exchange interactions at ambient conditions obtained from the converged GGA+U calculations.}
\vspace{0.1cm}
\begin{tabular}{| c | c | c  | c | c | c |}
\hline
  \hspace{0.03cm} $J_{ij}$  \hspace{0.03cm} & \hspace{0.03cm} Bond distances (\AA)  \hspace{0.03cm} & \hspace{0.03cm} Total (meV)  \hspace{0.03cm} & \hspace{0.03cm} $t_{2g}-t_{2g}$ (meV) \hspace{0.03cm} & \hspace{0.03cm}  $t_{2g}-e_{g}$ (meV) \hspace{0.03cm} & \hspace{0.03cm}  $e_{g}-e_{g}$ (meV) \hspace{0.03cm} \\
\hline
$J_1$ & 3.41 & -7.47 & -13.07 & 6.11 & -0.51 \\
\hline
 $J_2$  & 3.85 & 4.08 & -3.87 & 7.75 & 0.20 \\
\hline
\end{tabular}
\label{orbital}
\end{table*}
\egroup

\begin{figure}
\includegraphics[width=0.99\columnwidth]{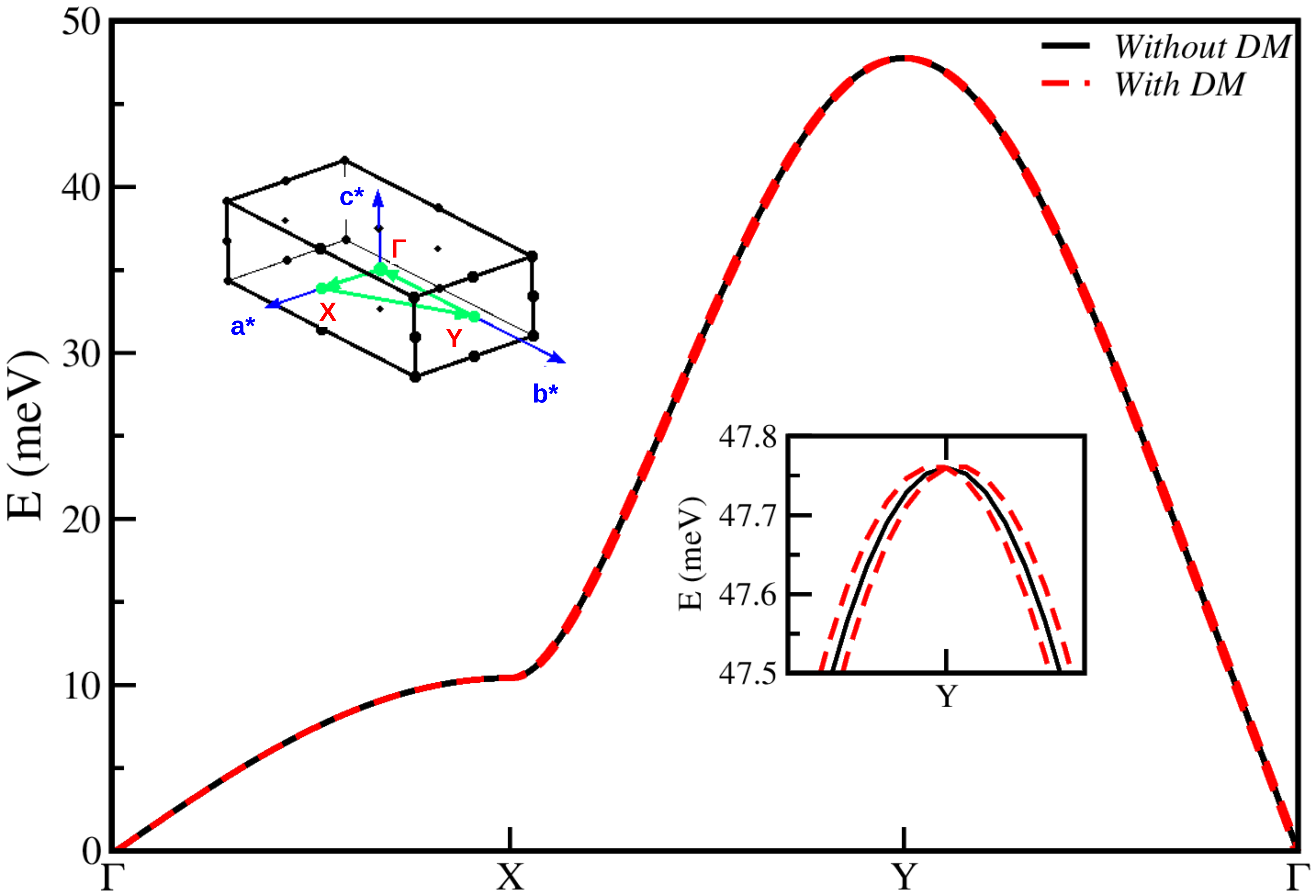}
\caption{The spin-wave dispersions along high-symmetry directions in the BZ computed with and without including DM-interactions. The inset shows the BZ and the labels of high-symmetry $k$-points. The inset also displays the spin-wave dispersions around Y-point within a smaller energy window to portray the degeneracy breaking due to the presence of DM-interaction.}
\label{spinwave}
\end{figure}

\par 
Aiming to provide a more complete microscopic physical picture and to assess the role of different orbitals in the resulting magnetism, we will now discuss the orbital decomposed $J_{ij}$ ($i$, and $j$ are two neighboring spin-sites, i.e. Cr in our case) in the crystal field basis. The  $e_g$ and $t_{2g}$-derived orbital contributions of dominant couplings such as $J_1$ and $J_2$ are shown in Table \ref{orbital}. We find that for both cases, the $t_{2g}$-$t_{2g}$ couplings are antiferromagnetic, $t_{2g}$-$e_{g}$ couplings are  ferromagnetic and $e_{g}$-$e_{g}$ couplings are negligibly small. Our results also reveal that the dominating contribution for  $J_1$ comes from $t_{2g}$-$t_{2g}$ coupling, while $J_2$ is dominated by $t_{2g}$-$e_{g}$ coupling. Another important point to note that $t_{2g}$-$t_{2g}$ contribution is much stronger for $J_1$, while $t_{2g}$-$e_{g}$ exchanges are almost equal for $J_1$ and $J_2$. This subtle difference in orbital dependency of magnetic coupling makes the overall nature of couplings different for 1st and 2nd NN Cr-ions. The nature and the relative magnitude of these orbital decomposed couplings could be physically understood within the framework of the extended Kugel-Khomskii model~\cite{Kugel1973,0038-5670-25-4-R03} . In this model, each Cr-Cr exchange coupling ($J_{ij}$) has ferromagnetic ($J_{ij}^{FM}$) and antiferromagnetic ($J_{ij}^{AFM}$) components as described below: 

\begin{equation} \label{eq1}
J_{ij}^{\mathrm{AFM}}={-4\sum_{t_{2g}-t_{2g}} \frac{\left(t_{t_{2g}-t_{2g}}\right)^{2}}{U} -4\sum_{t_{2g}-e_{g}} \frac{\left(t_{t_{2g}-e_{g}}\right)^{2}}{\left(U+\Delta_{t_{2g}-e_{g}}\right)}}
\end{equation}

\begin{equation} \label{eq2}
J_{ij}^{\mathrm{FM}}=4\sum_{t_{2g}-e_{g}}\frac{\left(t_{t_{2g}-e_{g}}\right)^{2} J_{H}}{\left(U+\Delta_{t_{t_{2g}-e_{g}}}-J_{H}\right)\left(U+\Delta_{{t_{2g}-e_{g}}}\right)}
\end{equation}

\begin{equation} \label{eq3}
J_{ij} = J_{ij}^{\mathrm{AFM}} + J_{ij}^{\mathrm{FM}}
\end{equation}

As evident from the above equations, the nature of the resulting inter-atomic magnetic interactions  ($J_{ij}$) depends on a few parameters, namely (a) crystal-field splitting ($\Delta_{t_{2g}-e_g}$), (b) effective virtual hopping strengths between the relevant orbitals ($t_{2g}-t_{2g}$, $t_{2g}-e_{g}$), (c) the nominal fillings of those orbitals, and (d) Hund's coupling strength ($J_H$) as well as the magnitude of Hubbard $U$ of correlated $d$-states. In our case, the relevant orbital for magnetism is half-filled Cr-$t_{2g}$ states and the empty $e_{g}$ states appear slightly above them due to the octahedral crystal field. The hopping being spin-conserved, up-spin electrons of a Cr$^{3+}$ are allowed to hopp to the neighboring Cr-site if the up-spin channel of that site is partially or fully empty. Since $t_{2g}$-states are exactly half-filled, inter-site $t_{2g}$-$t_{2g}$ virtual hoppings are allowed only if they possess anti-parallel alignments (see schematic in Fig.~\ref{schematic}(a) and (b)), making the $t_{2g}$-$t_{2g}$ exchange completely antiferromagnetic in nature and it has no contribution in $J_{ij}^{FM}$ as seen in equation~\ref{eq2}. Since both the spin-channels of $e_{g}$ states are empty, virtual inter-site $t_{2g}$-$e_{g}$ hoppings are allowed for both parallel and anti-parallel alignments (see schematic in Fig.~\ref{schematic} (c) and (d)). However, the parallel alignment is energetically favored because of the Hund's coupling and thus we see a ferromagnetic $t_{2g}$-$e_{g}$ coupling in our calculations. The magnitude of these couplings will primarily depend on the $t_{2g}$-$t_{2g}$ and $t_{2g}$-$e_{g}$ hopping strengths in an effective low energy Hamiltonian for $d$ states where the S-$p$ states are down-folded. In this picture, $e_{g}$-$e_{g}$ interaction should be zero. A small value obtained in our calculations arises due to the distortion in the octahedra, resulting in slight mixing of $t_{2g}$ and $e_{g}$-states. Thus, we provided the microscopic mechanism behind the nature of the orbital resolved couplings which appears a crucial quantity for explaining the resulting opposite sign of  $J_1$ and $J_2$, despite their similar bond distances. We believe that this mechanism is generic and should be useful in understanding the magnetism of other transition metal (TM) compounds where TM possess a nominal $d^3$ configuration (half filled $t_{2g}$). 

\begin{figure}
\includegraphics[width=0.99\columnwidth]{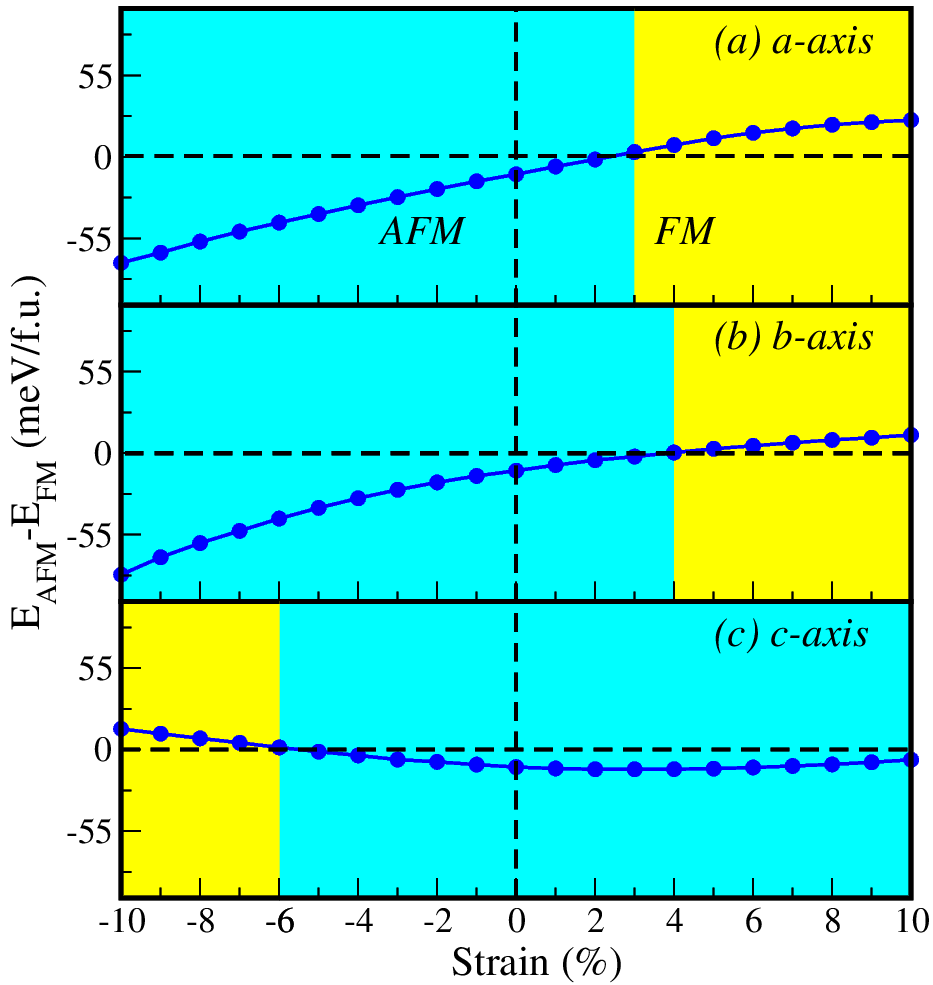}
\caption{The energy difference between FM and AFM states as a function of strain along (a) $a$, (b) $b$, and (c) $c$-axis is displayed. The magnetic transition is highlighted.}
\label{mag-strain}
\end{figure}
\par 
Next, we analyze the effect of relativistic spin-orbit coupling (SOC) which is known to play a crucial role in realizing novel exotic phases in many real materials. In the absence of SOC, the exchange interactions are isotropic with spin rotational invariance. However, SOC may lower the symmetry and gives rise to anisotropic interactions as theoretically explained by Moriya in Ref.~\cite{PhysRev.120.91} by extending the Anderson super-exchange theory~\cite{PhysRev.115.2}. The presence of spin-orbit coupling, leading to a sizable magnetocrystalline anisotropy (MAE), Dzyaloshinskii–Moriya (DM) interactions and symmetric anisotropic interactions has recently been studied in Cr-based layered compounds such as  CrX$_3$ (X=Cl, Br, I), CrGeTe$_3$ etc. The MAE helps to overcome thermal fluctuations and gives rise to long-range magnetic ordering in low-dimensional spin systems.  On the other hand, sizable DM interactions often compete with the Heisenberg $J_{ij}$ and give rise to topologically non-trivial spin textures.  We first estimated MAE by computing the energies corresponding to different spin quantization axis within the GGA+U approach incorporating SOC. The computed energies (Table \ref{MAE}) reveal that the crystallographic $a$-axis is the easy axis of magnetization (lowest energy). The spin moment on Cr-site is same for all quantization axes as expected and the orbital moment is 0.04 $\mu_B$ for the easy axis of magnetization. The MAE is defined as the energy difference between the easy and hard axis of magnetization. The computed value of MAE = $E_{1 0 0} - E_{0 1 0}$ = -0.38 meV.  Interestingly the magnetic moment favors aligning perpendicularly to the direction of the spin chain. The magnitude of the spin, orbital moments, and MAE in LaCrS$_3$ compare well with other reported Cr-based compounds such as CrI$_3$~\cite{PhysRevB.105.184430} where computed orbital moment and MAE are 0.07 $\mu_B$ and 0.3 meV, respectively. It is worthwhile to note here that despite having a small orbital moment, the effective SOC is strong in this compound as revealed from the large MAE. This could be attributed to the large spin-moment on Cr-site and the highly anisotropic crystal environment of this material which are essential requirements to exhibit large MAE in bulk materials. In few previous studies on Cr-based compounds~\cite{Lado_2017, Xu2018}, it has been claimed SOC is not directly contributed by the magnetic ion alone but induced by the non-magnetic ligand states. For instance, MAE is shown to arise due to the SOC of iodine ions~\cite{Lado_2017, Xu2018} and Te ions respectively~\cite{Xu2018} for CrI$_3$ and CrGeTe$_3$. In order to investigate this possibility, we have computed the site decomposed MAE as a function of SOC strength.  The contributions of La, Cr, and S-atoms to the total MAE are separately displayed in Fig~\ref{SOC}. In contrary to the previous reports~\cite{Lado_2017, Xu2018}, we found that contribution of La and S are negligibly small that the dominant contribution comes from the magnetic Cr-ions.

\begin{figure*}
\includegraphics[width=1.99\columnwidth]{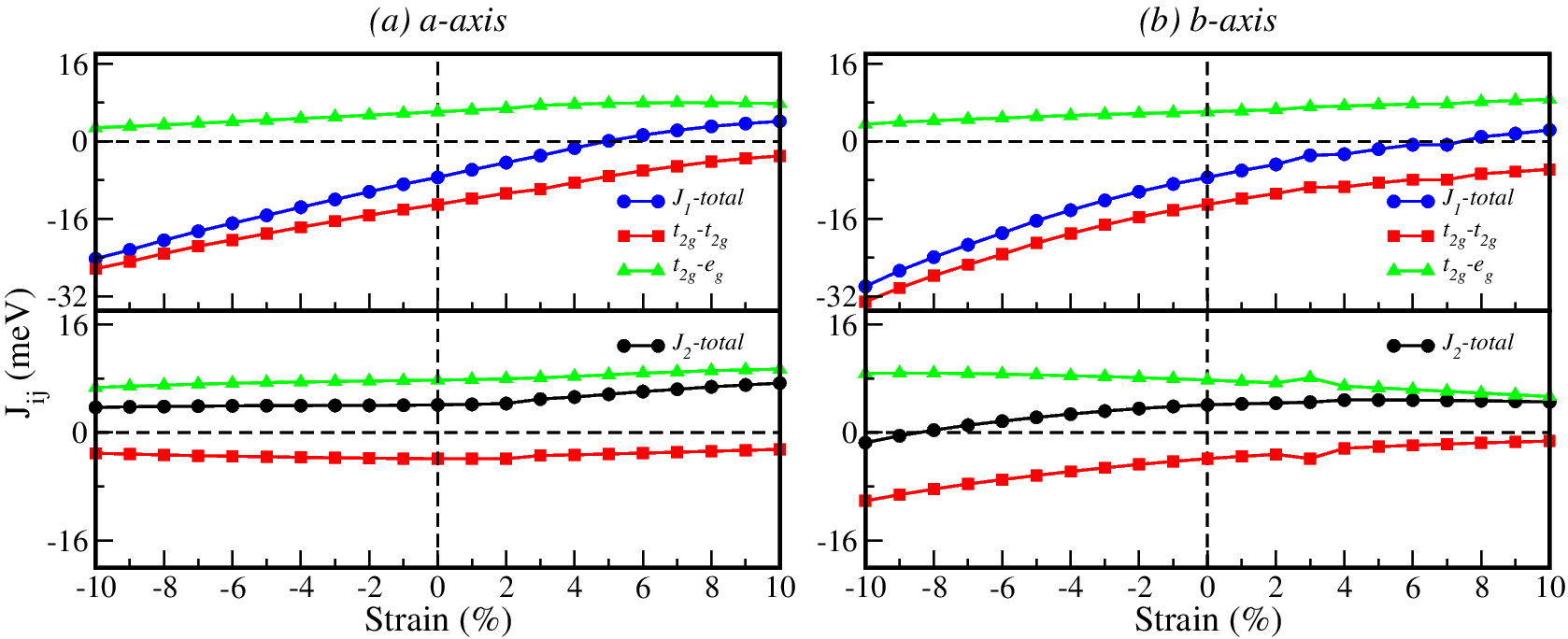}
\caption{The variations orbital resolved contributions ($t_{2g}$-$t_{2g}$, and $t_{2g}$-$e_g$) to the total $J_1$ and $J_2$ as a function of strain along (a) $a$ and (b) $b$-axis are displayed.}
\label{J-strain}
\end{figure*}


\begin{table*}[t]
\caption{The total energy, spin moments and orbital moments of Cr-ions for a given magnetization direction obtained from GGA+U+SOC calculations. The energies are relative to the energy of the cell with the magnetization along $b$ axis.}
\vspace{0.1cm}
\begin{tabular}{|   c    | c | c | c | c | c | c | }
\hline
 & \multicolumn{2}{c|}{$b$-axis} & \multicolumn{2}{c|}{$a$-axis} & \multicolumn{2}{c|}{$c$-axis} \\
 \hline
  & \hspace{0.1cm} Energy ($meV$) \hspace{0.1cm} & \hspace{0.03cm} Moment ($\mu_\text{B}$)  \hspace{0.03cm} &  \hspace{0.1cm} Energy ($meV$) \hspace{0.1cm} & \hspace{0.03cm} Moment ($\mu_\text{B}$)  \hspace{0.03cm} &  \hspace{0.1cm} Energy ($meV$) \hspace{0.1cm} & \hspace{0.03cm} Moment ($\mu_\text{B}$)  \hspace{0.03cm} \\
 \hline
 GGA+U+SOC & 0.00 & 2.89 (0.02) & -0.38 & 2.89 (0.04) & -0.25 & 2.89 (0.03)  \\
\hline
\end{tabular}
\label{MAE}
\end{table*}

Since the effective SOC is found significantly large and comparable with other Cr based materials, we further mapped the computed energies from GGA+U+SOC onto a generalized Heisenberg model as given below:  
\begin{equation}
    H=\sum_{i\neq j} e_{i}^\alpha J_{ij}^{\alpha \beta}e_{j}^{\beta}, \quad \alpha,\beta = x,y,z. 
\end{equation}

Here the magnetic exchange parameters $J_{ij}$ is a $3\times 3$ tensor for fully relativistic case and $\alpha, \beta$ represent tensor components along $x$, $y$ and $z$-axis. We calculated its all components to estimate DM interaction vector ($D_{ij}$) (antisymmetric anisotropic interaction) and symmetric anisotropic interaction ($C_{ij}$), which are given by,
\par
\begin{equation}
   D_{ij}^{x}= \frac{1}{2}\{J_{ij}^{yz}-J_{ij}^{zy}\},\quad D_{ij}^{y}= \frac{1}{2}\{J_{ij}^{xz}-J_{ij}^{zx}\},\quad D_{ij}^{z}= \frac{1}{2}\{J_{ij}^{xy}-J_{ij}^{yx}\}
\end{equation}
\begin{equation}
   C_{ij}^{x}= \frac{1}{2}\{J_{ij}^{yz}+J_{ij}^{zy}\},\quad C_{ij}^{y}= \frac{1}{2}\{J_{ij}^{xz}+J_{ij}^{zx}\},\quad C_{ij}^{z}= \frac{1}{2}\{J_{ij}^{xy}+J_{ij}^{yx}\}
\end{equation}
A detailed discussion on the implementation and calculations of these parameters is given in Ref.~\cite{PhysRevB.102.115162,PhysRevB.103.174422}. As shown in Table~\ref{Anisotropy}, the DM interaction for the first NN is zero due to the presence of space-inversion symmetry. However, the second NN interactions are relatively strong and the DM vector lies in $x$-$z$ plane which is perpendicular to the direction of the chain ($y$-axis). The magnitude of $C_{ij}$ is found to be very small for both the neighbors. Based on these interactions, we computed the spin-wave dispersions (see Fig.~\ref{spinwave}) using linear spin-wave theory as implemented in spin-w code~\cite{Toth_2015}. There are two Cr atoms which form the double chain in this system, resulting in two bands in the magnetic excitation spectrum. When anisotropic components of magnetic interactions are not considered, the spin wave dispersions are degenerate and the degeneracy is lifted after including DM interactions in the spin-Hamiltonian (see inset of Fig.~\ref{spinwave}). These results can be helpful for future neutron scattering experiments and analysis. We find that magnetic propagation vector $\nu$ = (0 0 0) comes out to be lowest in energy and the second nearest neighbor DM interaction does not yield any spin canting in this system.  We note that the nature of our computed spin-wave spectrum agrees well with the other spin-chain antiferromagnet such as RbFeS$_2$~\cite{PhysRevB.104.224419}. 
\begin{table}[t]
\caption{The three components of DM interactions vector and anisotropic symmetric exchange between 1st and 2nd NN Cr ions as obtained from GGA+U+SOC calculations. }
\vspace{0.1cm}
\begin{tabular}{|   c    | c | c | c | c  |c | c | c  | }
\hline
Site & Distance (\AA) &\hspace{0.1cm}{$D_x$}\hspace{0.1cm}& \hspace{0.1cm}{$D_y$}\hspace{0.1cm}& \hspace{0.1cm}{$D_z$}\hspace{0.1cm} & \hspace{0.1cm}{$C_x$}\hspace{0.1cm} & \hspace{0.1cm}{$C_y$}\hspace{0.1cm}& \hspace{0.1cm}{$C_z$}\hspace{0.1cm} \\
\hline
 1$^{st}$ NN& 3.41 & 0.00 & 0.00 & 0.00 & -0.002 &  0.002 & 0.006 \\
 \hline
 2$^{nd}$ NN &3.85 & 0.05 & 0.00 & -0.08 & 0.00  & -0.004 & 0.00 \\
\hline
\end{tabular}
\label{Anisotropy}
\end{table}
 
\subsection{Magnetic and electronic transition under uniaxial strain}
The strain engineering is considered to be a promising route for realizing electronic and magnetic transition in transition metal compounds~\cite{PhysRevLett.116.117202}. To explore possibilities like antiferromagnetic to ferromagnetic transition, tuning the effective SOC, switching the direction of easy/hard axis of magnetization, tuning the nature and magnitude of band-gap, etc, we have studied the magnetism and electronic structure of LaCrS$_3$ under the application of tensile and compressive strain along all the three crystallographic directions. The strain is defined as follows: $\nu = \frac{l-l_0}{l_0}$, where $l_0$ and $l$ are pristine and strained lattice constants, respectively. Thus the compressive strain corresponds to the negative values and positive values represent tensile strain. The value of $l_0$ has been taken from the experimentally reported lattice parameters~\cite{KIKKAWA1998233} and ionic relaxation has been performed for every value of $\nu$.  
\par 
\textbf{Lowest energy magnetic-state}:  
We have computed total energies corresponding to the FM ($E_{FM}$) and AFM state ($E_{AFM}$) by applying uniaxial strain along all the three directions and the energy difference $\delta E = E_{AFM}- E_{FM}$ presented in Fig.~\ref{mag-strain}(a)-(c). We found that due to compressive strain along $a$ and $b$-axis, $\delta E$ increases monotonously along the negative direction, indicating stronger stability for the AFM state. However, $\delta E$  becomes zero at about $2\%$ of tensile strain and moves towards positive values upon enhancing the tensile strain. Thus, our results reveal a magnetic transition from AFM state to FM state just by applying  $2\%$ of tensile strain. However, the change of $\delta E$  is much smaller when the strain is applied along the $c$-direction. This is not surprising considering the fact that Cr-ions are located in the $a-b$ plane and well separated along the $c$-directions (5th NN). Hence all the remaining discussions are focused on the strain along $a$- and $b$-directions. 

\begin{figure*}
\includegraphics[width=1.99\columnwidth]{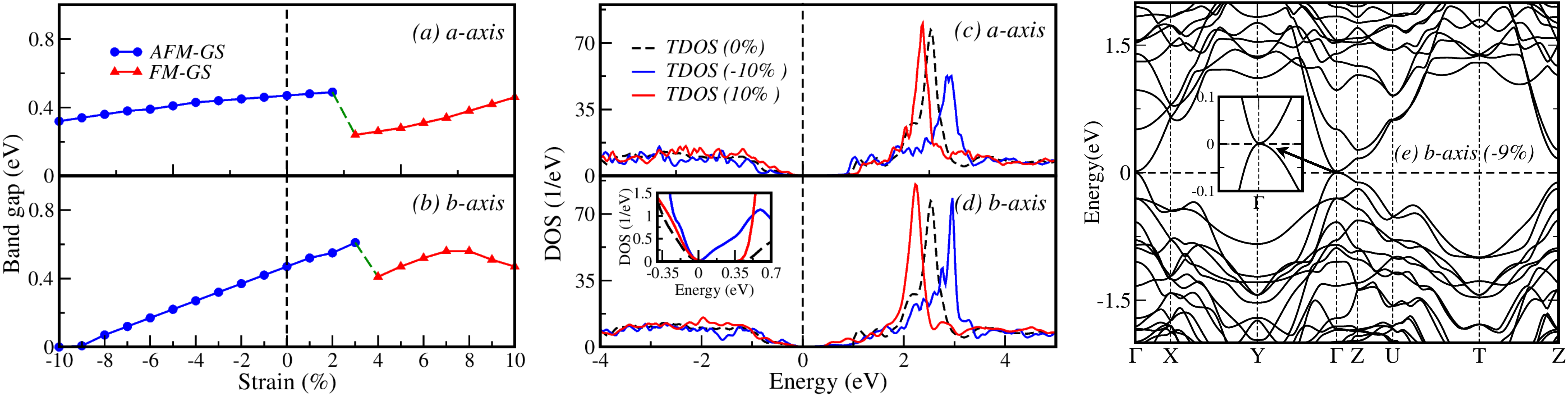}
\caption{The variation of the band-gap in the AFM and FM ground state (GS) as a function of strain along (a) $a$, and (b) $b$-axis. The total DOS  in their respective magnetic ground state for three values of strains (+10$\%$,  0$\%$, and -10$\%$ ) along (c) $a$, and (d) $b$-axis. Inset in (d) shows TDOS in the close vicinity of Fermi energy to portray vanishing energy gap at -10$\%$. (e) The band dispersion at the critical value of strain along $b$-axis (-9$\%$) where band-gap becomes zero.  It clearly displays the realization of an interesting non-gap semi-conducting state where band gap is zero only at the centre of the BZ ($\Gamma$-point). Inset demonstrates the dispersion at and around $\Gamma$-point.}
\label{gap-dos}
\end{figure*}

\textbf{Microscopic origin of magnetic transition}:  
In order to understand the origin of magnetic transition, we computed the dominant magnetic exchange interactions ($J_1$, $J_2$)) and their orbital decomposition as a function of external uniaxial strain. The changing trend of both couplings for strain along $a$ and $b$ axes are displayed in Fig.~\ref{J-strain}(a) and (b), respectively. The results show that the compressive strain region is strongly dominated by antiferromagnetic $J_1$, while in the tensile regions, $J_1$ and $J_2$ have comparable magnitudes. A critical value of tensile strain is able to make the net coupling ($J_{tot}$ = $J_1$+$J_2$) positive, giving rise to the occurrence of magnetic transition as discussed above. Our primary observations can be summarized as follows: (i) the first nearest neighbor exchange coupling ($J_1$) strongly varies under the application of strain along both axes ($a$ and $b$), (ii) the influence of strain on the second NN coupling $J_2$ is minimal, (iii) the orbital resolved exchange couplings for both uniaxial strain reveal that $t_{2g}$-$t_{2g}$ coupling is responsible for the strong variation of $J_1$, while $t_{2g}$-$e_{g}$ coupling does not vary significantly as a function of strain, (iv) in case of $J_2$, the ferromagnetic component, $t_{2g}$-$e_{g}$ coupling always remain slightly stronger than antiferromagnetic $t_{2g}$-$t_{2g}$ coupling and their magnitudes are small, making the net coupling weak ferromagnetic, (v) interestingly in the entire region of our study, the nature of $t_{2g}$-$t_{2g}$ coupling remain antiferromagnetic and $t_{2g}$-$e_{g}$ is ferromagnetic which is consistent with Kugel-Khomskii model as discussed before (see eqn~\ref{eq1} and ~\ref{eq2} and the corresponding analysis). Thus our results explain the microscopic origin of the strongly antiferromagnetic spin chain region under the influence of compressive strain and the relatively weaker ferromagnetic spin chain region after a critical value of tensile strain. We also note that at higher values of compression, $J_1$ is significantly stronger than $J_2$ and hence the system will behave like a uniform antiferromagnetic spin chain compound. 
\par 
\textbf{Effect of strain on electronic structure}:  
To understand the effect of strain on the electronic properties, we computed the band gap for various values of strain along $a$ and $b$-axis. The results are displayed in Fig.~\ref{gap-dos}(a) and (b) which illustrate that the system remain insulator even after the magnetic transition to FM state under the influence of tensile strain. However, the band gap shrinks compared to the ambient condition. Most interestingly, we could realize a FM insulating state in the tensile region. On the other hand, we observe a monotonic reduction of band-gap as compressive strain enhances and at $-9\%$ strain along $b$-axis, it becomes zero. The calculated total DOS for strain along $a$ and $b$-axis are displayed in Fig.~\ref{gap-dos}(c) and (d), respectively. These results indicate that valence band does not significantly change  due to strain, while the conduction band arising from Cr-$d$ states broaden due to compression and a small spectral weight at Fermi level is observed for -10$\%$ strain along $b$-axis (see inset of Fig.~\ref{gap-dos}(d)). The band dispersion along the high-symmetry paths of the Brillouin zone at the critical value of strain (-9$\%$) is displayed in  Fig.~\ref{gap-dos}(e). Remarkably the band-dispersion has a very interesting character that the valence band and conduction band edge touches each other near $\Gamma$-point and the gap still persists at all other points. These results indicate that compressed LaCrS$_3$ exhibits a novel gapless semiconducting antiferromagnetic ground state which has potential applications in spintronic devices~\cite{gaplesssemiconduct-book,Wang2010,PhysRevLett.100.156404}. The gapless semiconductor is a very interesting class of materials where threshold energy is not required to move electrons from valence to conduction band and thus the electron mobility is much higher than the usual semiconductors. The most famous example in this category is graphene which has gapless band structure and linear energy-momentum dispersion~\cite{RevModPhys.81.109}. There are few other examples also such as PbPdO$_2$~\cite{PhysRevLett.100.156404}, and VTaNbAl~\cite{PhysRevB.107.134434}. However, none of these materials are antiferromagnetic. The antiferromagnetism along with gapless electronic structure is truly rare. Thus, our results provide a route to obtain such a novel ground state in a transition metal compound. 
\begin{figure*}
\includegraphics[width=1.99\columnwidth]{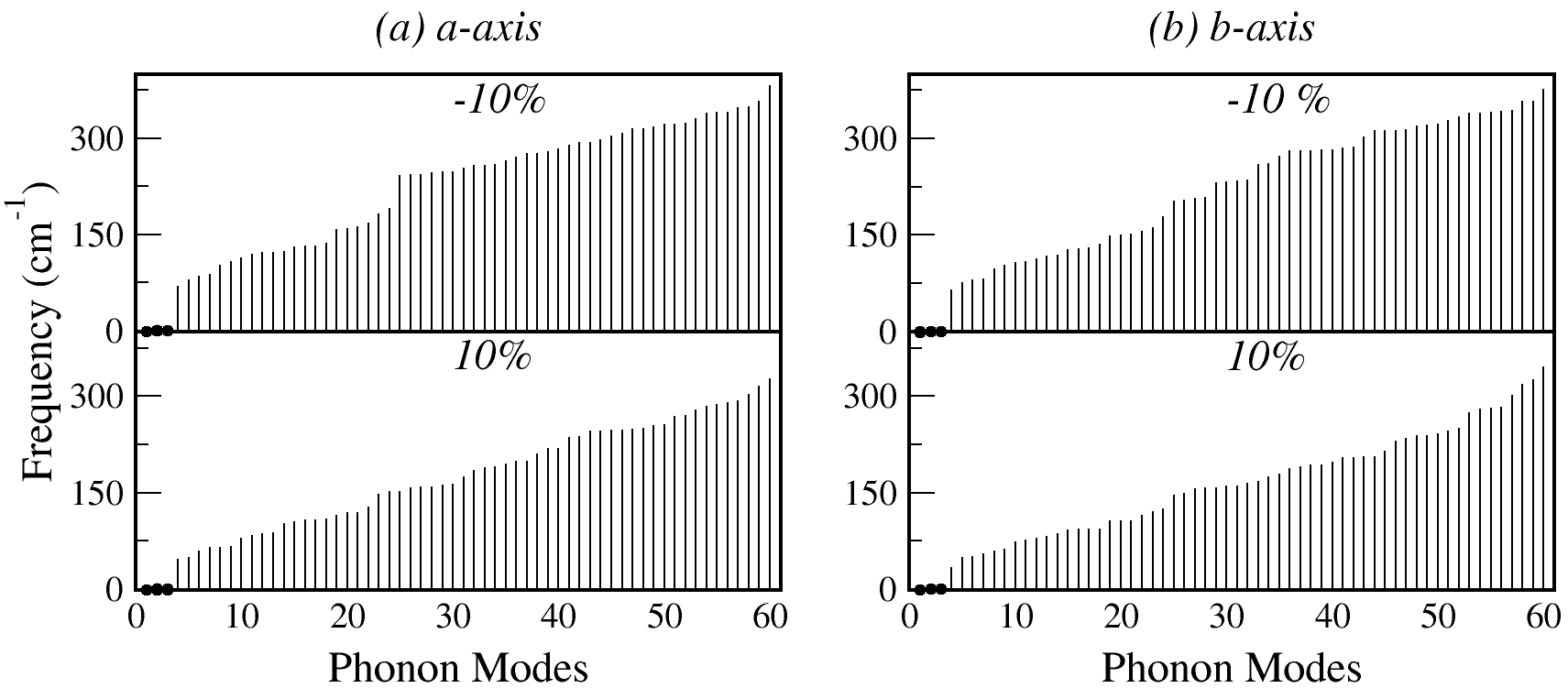}
\caption{Phonon frequencies (in $cm^{-1}$) at $\Gamma$-point of the Brillouin zone under (a) compressive strain, and (b) tensile strain. The mode numbers from 1 to 60 (total 60 modes corresponding to 20 atoms in the unit cell) is displayed in $x$-axis and the bar in $y$-axis denotes the corresponding frequency. We note that the first few frequencies are very small for all cases.}  
\label{Phonon}
\end{figure*}
\subsection{Stability Analysis}
To account for the stability of the system under compressive and tensile strain, we have calculated the phonon frequencies at $\Gamma$-point of the Brillouin zone from the converged GGA+U solutions. The finite difference method as implemented in VASP code has been employed. In this approach, the second order derivative of the total energy with respect to ionic displacements i.e. the force constants or Hessian matrix are calculated using finite differences. The dynamical matrix is thus constructed from force constants and diagonalized to obtain frequencies corresponding to each phonon mode. The calculated phonon frequencies under strain along $a$ and $b$ axes are shown in Fig. \ref{Phonon}(a) and (b), respectively. It can be seen that there are total 60 frequencies corresponding to the twenty atoms in the unit cell. Our results find that all the modes under 10\% compression and stretch are positive. Since there are no imaginary frequencies, the system is considered to be stable under strain.
\section{Conclusion}
In conclusion, we studied the electronic and magnetic properties of LaCrS$_3$ using first-principles DFT+U calculations where Hubbard $U$ was estimated from cRPA method. We found that the system is antiferromagnetic and semi-conducting  with a direct band-gap of 0.5 eV at ambient condition. Our computed exchange interactions based on magnetic force theorem reveal that Cr-spins form a non-frustrated chain network which propagates along crystallographic $b$-axis. Within this chain, the first NN interaction is found to be strongly antiferromagnetic, while second NN is weak ferromagnetic. We showed that the contradictory nature of these inter-atomic (Cr-Cr) magnetic interactions despite very similar bond-distances could be understood based on orbital resolved components in crystal-field basis. Our calculations demonstrate that SOC effect is significant, leading to sizable values of MAE and DM interactions. The strengths of these quantities are comparable with other Cr-based compounds such as CrI$_3$. Further, we applied mechanical strain along $a$ and $b$-axis and found that system exhibit magnetic transition to an insulating FM ground state due to tensile strain. On the other side, the compressive strain gives is able to strongly enhance the first NN interaction, giving rise to a uniform antiferromagnetic spin chain behavior in LaCrS$_3$. Interestingly a gap-less AFM ground state could be also realized for -9$\%$ strain along $b$-axis where the valence band and conduction just touch each other near $\Gamma$-point and all the other parts of the BZ are gaped. Thus, under the application of strain, we report two interesting transitions, namely AFM to FM transition and normal semiconductor to a gapless semiconductor transition in LaCrS$_3$. The realization of such gapless semi-conducting state in LaCrS$_3$ opens up huge possibilities for industrial applications in making new spintronic devices.

\end{document}